\documentclass{birkjour}
\usepackage{comment}

 \theoremstyle{definition}
 
 \theoremstyle{remark}

 \numberwithin{equation}{section}
 
\usepackage{graphicx}
\usepackage{amsmath}
\usepackage{float}
\usepackage{amsfonts}
\usepackage{bbold}

\usepackage{physics,braket,bm,amssymb} 
\newcommand{\pa}[2]{\dfrac{\partial #1}{\partial #2}} 

\newcommand{\p}[1]{\left(#1\right)} 

\newcommand{\sbr}[1]{\left[#1\right]} 

\newcommand{\f}[2]{\dfrac{#1}{#2}}

\date{\today}
\begin{document}

%
%
%
%
%
%
%
%
%

\title {Theory of Irrotational Flow}

\author{Steven Nerney}

\address{Independent Scholar, Boulder, Colorado, USA}

\email{*sfnerney@gmail.com}

\author{Edward G. Nerney}
\address{Laboratory for Atmospheric and Space Physics, University of Colorado Boulder, Boulder, Colorado, USA}
\email{Edward.Nerney@lasp.colorado.edu}

\begin{abstract}
We analyze the Navier Stokes equation, and show that all non-viscous, irrotational flows are barotropic. As far as we know, this has never been stated in the literature before, and indirectly
suggests that vorticity is required to make these flows non-barotropic. Both the pressure, temperature, speed, entropy, enthalpy, and a  Bernoulli function are derived to be only functions of the density (which is a function of space and time). 
When we also require steady flow, the Bernoulli function is a constant in all space, not just on streamlines. 
\end{abstract}

\maketitle

\section{\label{sec:level1}Introduction}

Unsteady fluid flow problems in 3D with viscosity and heat transfer involve mathematical and numerical difficulties so great that we must look to simplifications of the Navier-Stokes equations to make progress. It is necessary to derive relatively simple models that can be treated analytically. Hopefully, we will then learn something concerning the underlying more complex flows \cite{ref1}.

In this manner, we study irrotational flow where we ignore vorticity in the flows surrounding a thin layer near a boundary. Irrotational solutions predict pressure forces on a stream-lined body that agree well with experimental data as the viscosity becomes small \cite{ref2}. Further, the boundary layer equations show that the pressure is approximately constant across the layer. Then the pressure throughout the layer is equal to the pressure on the surface of the boundary and can be found from a solution of the irrotational flow around the body \cite{ref2}.  

When shocks are present, the boundary layer can strongly influence the flow outside the boundary. Large pressure gradients in the shock can make those gradients dependent on the effects they produce in the boundary layer \cite{ref1}. Here we examine irrotational flows without shocks and assume that the non-slip condition enforced by the boundary layer has a negligible effect on the surrounding flow pattern. 

We discuss theorems related to but more general than Kelvin's circulation theorem. It is important, then, to discuss the differences between our work and Kelvin's work.
Kelvin's theorem shows when a given irrotational flow will remain irrotational. The theorem requires that there is no net viscous force and the body forces must be conservative. 
The fluid pressure must be barotropic, $ P=P(\rho )$,  the reference frame must be inertial, and the fluid must be Newtonian \cite{ref2}. 

This study follows a different approach than discussed in Kelvin's theorem, but we do require a Newtonian fluid in an inertial frame of reference, with a constant coefficient of viscosity. Given only the irrotational flow assumption for all time, we ask what this implies for the general 
solutions of the Navier-Stokes equations. We include unsteady, 3D flow, no assumptions for P or $\rho $, and an irrotational viscous force (which is later set to zero to derive barotropic flow). The inclusion of arbitrary compressibility implies that Kelvin's theorem doesn't apply.

It is quite surprising that the non-viscous, 
irrotational Navier-Stokes equations require that the fluid must be barotropic, $ P=P(\rho)$. This is not assumed but is derived based {\it only } on the assumption of non-viscous, irrotational flow. 
Temperature is then derived from an equation of state (the ideal gas law) which  shows that $T=T(\rho) $. The first law of thermodynamics is shown to imply that s (entropy) and h (enthalpy) are only functions of density. 

The reader is asked to visualize a surface of constant density undulating in time and, everywhere on the 3D surface, P, $\rho $, T,  v (speed), s, h, and the Bernoulli function are different constants for non-viscous flow. Even so, the four quantities may be quite different functions of space and time because the density is $\rho =\rho (\mathbf r, t)$ and the variables are different functions of $\rho $. 

As an aside, the closure problem for the non-viscous  Navier-Stokes equations is thus circumvented because the fluid is barotropic, and conservation of mass and momentum give four equations in four unknowns $(\mathbf v, \rho )$, which are then reduced using  $v=\bm \nabla \Phi$ to two equations in two unknowns for $\Phi , \rho $.

It is well-known that Bernoulli's function is constant along streamlines or on vorticity lines for non-viscous, steady, barotropic flow. What we show here is that the   Bernoulli function must be only a function of $\rho (\mathbf r, t)$, even for unsteady, viscous flow. Further assuming steady, viscous flow, the Bernoulli function is a constant over all space, not just on streamlines. 

\section{\label{sec:level2}Unsteady, Compressible, Viscous Flow}

{\it Theorem 1: Assuming unsteady, compressible, viscous, irrotational flow, the irrotational viscosity is absorbed into the gas pressure, which is called $P_{\mu }$ (see 2.7). $P_{\mu }$ and the Bernoulli function are only a function of $\rho (\mathbf{r}, t)$ . Barotropic flow further requires $v=v(\rho) $.}

We make the following assumptions for irrotational flow: a Newtonian fluid, an inertial frame of reference, and a constant coefficient of viscosity (We point out that real Newtonian fluids have temperature dependent transport coefficients ; see figure A-3 in White \cite{ref3} ). 

We begin with the continuity equation: 

\begin{align}   \label{eq:1}  
   \pa{\rho}{t}+\bm\nabla \cdot \p{\rho \mathbf v}=0 
\end{align}

The momentum equation is written as:
\begin{align} \label{eq:2}  
\pa {\mathbf v}{t} +\mathbf v\cdot \bm\nabla \mathbf v=-\frac {\bm\nabla P}{\rho }-\mathbf \bm\nabla \Omega
+\frac {\mu } {\rho } \nabla ^2 \mathbf v
\end{align}

where $\mathbf v, P, \rho ,\Omega, \mu$ are  the velocity, pressure, mass density, an arbitrary conservative potential energy/mass,  and coefficient of viscosity, respectively. The arbitrary conservative force/mass is $\mathbf f=-\bm\nabla\Omega$.

We use the following well-known identities to simplify the momentum equation, (2.2),  and define the Bernoulli function, with the vorticity defined as  $\bm \omega \equiv \bm\nabla\times\mathbf v$
\begin{align}
\mathbf v\cdot \bm\nabla \mathbf v=\bm \omega \times \mathbf v+\bm\nabla v^2/2   \label{eq:3}   \\
\nabla ^2 \mathbf v = \bm\nabla (\bm\nabla \cdot \mathbf v)-  \bm\nabla \times \bm \omega  \label{eq:4} 
\end{align}

Using the identities, (2.3, 2.4), and substituting into (2.2), as well as absorbing the irrotational viscous force into the pressure:
\begin{align}
\pa {\mathbf v}{t}+\bm \omega \times \mathbf v=-\dfrac {1} {\rho} \bm\nabla (P- \mu \bm\nabla \cdot \mathbf v ) -\bm\nabla Be - \frac {\mu } {\rho } \bm\nabla \times \bm \omega  \label{eq:5} \\ 
Be\equiv  \frac {v^2} {2} +\Omega \label{eq:6}
\end{align}

Now we set $\bm \omega =0 $ for irrotational flow so we can write $\mathbf v=\bm\nabla \Phi $. 
 We define $P_{\mu }$ and rewrite the momentum equation as :

\begin{align} \label{eq:7}  
P_{\mu }\equiv P- \mu \bm\nabla ^2 \Phi \\
Be \equiv  \pa {\Phi} {t}  + \frac {1} {2} \bm\nabla \Phi \cdot \bm\nabla \Phi +\Omega \\
0= -\frac {1} {\rho }\bm\nabla P_\mu  - \bm \nabla Be
\end{align}

We now take the curl of (2.9) to find

\begin{align} \label{eq:8}  
\bm\nabla \times \left( \frac {\bm\nabla P_{\mu }}{\rho } \right)= -\frac {\bm\nabla \rho \times \bm\nabla P_{\mu } } {\rho ^2} =0
\end{align}

This requires that $\bm\nabla P_{\mu }$ and $ \bm\nabla \rho $ are locally parallel and indirectly requires that $P_{\mu }=P_{\mu }(\rho )$. 
This may be visualized by picturing a 3D surface of constant $\rho $. $\bm\nabla \rho $ is perpendicular to this surface  and so is $\bm\nabla P_\mu  $ . 
Then the surface of constant density is also a surface of constant $P_{\mu }$. 
We label the surface values $(\rho _1, P_ {\mu 1})$ and then imagine a slightly different surface with values $(\rho _2, P_{\mu 2})$. The ith surface has $(\rho _i, P_{\mu i} )$. 
This list of values defines the functional dependence of $P_{\mu } (\rho )$ on $\rho $; i.e., it determines $P_{\mu }=P_{\mu } (\rho )$. 
We see that the irrotational flow assumption forces $P_{\mu }=P_{\mu }(\rho)  $ .  

The case where $\rho \to 0$ in (2.10) is evaluated using L'Hospital's rule and it 
can be shown that  
$\bm\nabla\f{\partial^2 P_{\mu }}{\partial\rho^2} \times \bm\nabla \rho=0$  so we require  $\f{\partial^2 P_{\mu }}{\partial\rho^2}$ to be a function of $
\rho$ in the limit $\rho\to0$. 

Now cross $\bm \nabla \rho $ into (2.9). 

The first term is zero so we know that $ \bm\nabla \rho  \times\bm\nabla Be = 0$. By the identical 
 logic as before, $Be=Be\p{\rho}$. 
 
It is often the case that the Navier-Stokes equations are solved for either isothermal or adiabatic flow. Then (2.7) indicates that $\bm \nabla ^2 \Phi $ must be a function of $\rho $. Steady flow in (2.1) allows the continuity equation  to be written as:
\begin{align} \label{eq:9}  
\bm \nabla ^2 \Phi = -\frac {1} {\rho } \bm v\cdot \bm \nabla \rho 
\end{align}
Because the left-hand-side is a function of $\rho $, the speed also must be a function of $\rho $. We note that this is for steady, viscous, barotropic  flow.
\section{\label{sec:level3}Unsteady, Non-Viscous, Compressible Flow}

{\it Theorem 2: Irrotational, compressible, non-viscous,  unsteady flows are barotropic; i.e. $P=P(\rho )$}.

Setting $\mu=0$ for the moment, we see that $P_\mu =P$, so that P must be $P=P(\rho) $. We have not assumed barotropic flow, but we see that it is required for irrotational, compressible, non-viscous, 3D, unsteady flow. 
Notice that these general flows require that the surfaces of constant $P, \rho , Be $ all move as one (due to the unsteady flow) and are all only a function of $\rho $, even though each will have different functional dependencies on time and space through $\rho ( \bm r,t)$. 


\section{\label{sec:level4}Steady, Viscous,  Irrotational Flow}

{\it Theorem 3: Steady, viscous, compressible, irrotational flow requires that the Bernoulli function is constant in all space. Setting the generalized conservative force, $\Omega $}, to zero, requires $v =v (\mathbf \rho )$

We now examine steady flow.
The momentum equation, (2.9), can now be simplified, although we must further assume a steady state. Derivations of the  equation (4.1) use $dP_{\mu }= \bm dr \cdot \bm \nabla P_{\mu }$. There is an extra term with time dependence, $ dt \pa {P_{\mu }} {t}$, so we must assume a steady state to go further.
Steady flow allows the absorption of $\dfrac {\bm\nabla P_{\mu }} {\rho} $ into the Bernoulli function as
 
\begin{align} \label{eq:10}  
\frac  { \bm\nabla P_{\mu } }  {\rho } =\bm\nabla \int \frac  { dP_{\mu }}  {\rho } 
\end{align}

and is derived from 

\begin{align} \label{eq:11}  
\mathbf dr \cdot \left  (\frac  { \bm\nabla P_{\mu } }  {\rho }\right ) = d\int \frac {dP_{\mu }} {\rho } = \mathbf dr \cdot \bm\nabla \int \frac  { dP_{\mu } }  {\rho } 
\end{align}

We note that this derivation is only valid for $P_\mu =P_\mu (\rho ) $. This can be seen by taking the curl of (4.1) which re-derives (2.10). 

The $P_{\mu }$ gradient term is now absorbed into the Bernoulli equation using (4.1) so that (2.9) becomes: 

\begin{align} \label{eq:12}  
 \bm\nabla \left( \int \frac {dP_{\mu }} {\rho } +\frac {1} {2} \bm\nabla \Phi \cdot \bm\nabla \Phi +\Omega  \right) = 0
\end{align}

This means that the Bernoulli function is a constant in all space, which we call E, the total energy/mass in the flow. 

\begin{align} \label{eq:13}  
 E=  \int \frac {dP_{\mu }} {\rho } +\frac {1} {2} \bm\nabla \Phi \cdot \bm\nabla \Phi +\Omega
\end{align}
When $\Omega =0$, $v=v(\rho )$ because $P_\mu = P_\mu (\rho )$. 

Curiously, we know that $P_\mu $ must be $P_\mu(\rho)$ so that $\int \frac {dP_\mu (\rho ) } {\rho } $ can be done in principle. Yet we have no knowledge of $P_\mu(\rho)$ without further assumptions.

We comment on Beltrami flows which have $\bm\omega ,\mathbf v$ parallel to each other so that $ \bm \omega \times \mathbf v$ is zero but $\bm\omega $ is nonzero. The momentum equation, (2.2), can be simplified to (2.9), but only with {\it steady, non-viscous} Beltrami flow. Then the  derivation in section 3 is still correct with $P(\rho )$, but the assumptions break down. Beltrami flows obey the vorticity equation, $\pa {\bm \omega } {t} = \mu  \nabla ^2 \frac {\bm \omega } {\rho} $ so that some driving term is likely necessary
to avoid the solution decaying to zero vorticity. 

 \section{\label{sec:level5}Steady, Non-Viscous, Compressible, Irrotational Flow}

{\it Theorem 4: Steady, non-viscous, compressible,  irrotational flow 
allows the pressure integral, $\int \frac {dP(\rho )} {\rho } $, to be integrated and shows that the speed, v, is $ v=v(\rho )$. }

In this case (4.4) becomes
\begin{align} \label{eq:16}  
 E=  \int \frac {dP(\rho )} {\rho } +\frac {1} {2} \bm\nabla \Phi \cdot \bm\nabla \Phi +\Omega
\end{align}

All steady, non-viscous,  irrotational flow problems come down to solving (5.1) together with the continuity equation. The differences are in the boundary conditions and the pressure integral. 

Having shown that $P=P(\rho )$ in section 3, we use the following particular form for $P(\rho )$:

\begin{align} \label{eq:15}  
P=k_1\rho ^{\gamma }\\
k_1 = \frac {P_0 } {\rho _0^{\gamma } }
\end{align}

where $\gamma =5/3, 1$ for either an adiabatic or isothermal, monatomic ideal gas. We also include the case where $\gamma $ is some other constant. 
This form has been used in studying the solar wind (the Sun's outer atmosphere) where the wind is heated in depth by an unknown process. Then gamma approximates the unknown heating function \cite{ref4} .

\begin{align} \label{eq:17}  
 \int \frac  { dP }  {\rho } = \int k_1 \gamma \rho ^{\gamma-2} d\rho =\frac  {\gamma } {\gamma -1}  k_1\rho ^{\gamma -1}
\end{align}
for the adiabatic case or the general case for $\gamma \ne 1$,and

\begin{align} \label{eq:18}  
 \int  \frac  {dP}  {\rho } = \int \frac {kT}{m\rho}  d\rho =\frac {kT} {m}  \ln\rho
\end{align}

for the isothermal case with $\gamma=1$. The ideal gas law has also been used: 

\begin{align} \label{eq:19}  
P= nkT=\frac {kT} {m} \rho 
\end{align}
where k, m, and n are Boltzmann's constant, the average mass/molecule, and the number density, respectively.

Each form for the pressure integral (internal energy/mass), (5.4, 5.5), are to be substituted into (5.1) for a particular problem.
Setting $\Omega=0$, (5.1) indicates that $ v=v(\rho )$. 
Currie \cite{ref5} also shows this in a homework problem for 2D flow, while our result is for 3D flow.   

\section{\label{sec:level6}Unsteady, Irrotational, Non-Viscous Entropy Changes}

{\it Theorem 5: The unsteady, irrotational, non-viscous, compressible temperature, T,  entropy, s, and the enthalpy, h, are functions of $\rho (\mathbf{r}, t) $ .}

Crocco's equation is often used to show that isentropic flow occurs if and only if the flow is irrotational when the stagnation enthalpy is constant everywhere \cite{ref5}. 

Here we have unsteady, compressible flow with a variable stagnation entropy so that the  assumptions of Crocco's theorem are violated; but we follow the same basic logic to find the limitations for irrotational flow when  including heat flow. 
Non-viscous flow means that the fluid is barotropic as shown in section 3.
Using the first law of thermodynamics

\begin{align} \label{eq:22}  
de= -Pd \left (\frac {1} {\rho } \right) +Tds
\end{align}

where e is the internal energy/mass,  T is temperature, and s is the entropy/mass. 
The enthalpy/mass is defined as

\begin{align} \label{eq:23}  
h=e+P/\rho 
\end{align}
 
Because the internal energy of an ideal gas only depends on T in equations  (6.1, 6.2), together with $T=T(\rho )$ (from the ideal gas law), 
we can solve for ds and integrate term-by-term:

\begin{align} \label{eq:24}  
\Delta s= \int \frac {P(\rho )} {T(\rho)} d \frac {1} {\rho }  +\int \frac {d \hspace{.04cm} e(T(\rho ))} {T(\rho )}  
\end{align}
This  shows that barotropic flows must have $s=s(\rho ), h=h(\rho )$, and also means that $\bm \nabla s, \bm \nabla T$ are always parallel.
 
An alternative way to show this for steady flow uses the vector form for the total differential.

 We eliminate e using h in (6.1): 
 
 \begin{align} \label{eq:25}  
dh - d \left (\frac {P} {\rho } \right )= -Pd \left (\frac {1} {\rho } \right )+T ds
\end{align}

 and use  $d\mathbf r \cdot \bm\nabla =d $ to find
 
 \begin{align} \label{eq:26}  
 T\bm\nabla s = \bm\nabla \left ( h - \int \frac {dP} {\rho}  \right )
\end{align}
 
 We take the curl of both sides 

 \begin{align} \label{eq:27}  
\bm\nabla T\times \bm\nabla s = 0
\end{align}

 Again this implies that s is a function only of $T(\rho ) $ by the same logic in section 2.

\section{Steady, Non-Viscous, 2D, Irrotational Flow}

{\it Theorem 7: Steady, non-viscous, compressible, 2D, irrotational flows have a vector potential with one component, which gives the equation for the streamlines. The coordinate  normal to the streamlines, $\eta $, allows the calculation of the percentage change in the kinetic energy/mass along $\eta $, which  is proportional to the average curvature of the streamlines over $\Delta \eta $. }

We previously proved that $P=P(\rho )$ for non-viscous flow (section 3). We also proved that the Bernoulli function is constant in all space for steady flow (not just on streamlines) in section 4. A further restriction to 2D flow shows that the solution can be constructed from the single component of the vector potential for the steady continuity equation.
The steady version of equation (2.1) may be written as:

\begin{align} \label{eq:34}  
 \rho \mathbf{v} = \bm \nabla \times \mathbf{A}
\end{align}

Because the velocity only has two components (say $v_r,v_\theta$) then we write out the components of (7.1):

\begin{align} \label{eq:28}  
  \mathbf{A} = A(r, \theta ) \hat e_z\\
  r\rho \pa {\Phi} {r} = \pa {A} {\theta } \\
  \frac{\rho } {r} \pa  {\Phi} {\theta } = -\pa {A} {r}
\end{align}

This is equivalent to using a Stokes function. 

Then it follows by direct substitution that $\mathbf{v} \cdot \mathbf \nabla A=0 $. This means that the function $A(r, \theta )$ is constant on streamlines so that $ A(r, \theta )$  is the equation for the streamlines. 

Setting $\Omega =0$, the simplified momentum equation may be written as:

\begin{align} \label{eq:29}  
  \mathbf{v} \cdot \mathbf{\nabla} (v\hat e_s ) = -\mathbf{\nabla} \int \frac {dP} {\rho } 
\end{align}
where $\hat e_s$ is the unit tangent vector to the streamlines and v is the speed. 

The component of $\nabla $ along  streamlines gives the integrated Bernoulli equation, while the component perpendicular to streamlines gives:

\begin{align} \label{eq:36}  
 v^2  \pa {\hat e_s} {s}  = -\pa {\left( \int \frac {dP} {\rho } \right)} {\eta }
\end{align}

where $\eta $ is the coordinate perpendicular to streamlines. Eliminating the pressure integral by using (5.1):

\begin{align} \label{eq:37}  
 -\frac {v^2} {R} = \pa {\left( \frac {v^2} {2} \right)} {\eta } 
 \end{align}
 
 where R is the instantaneous radius of curvature of the streamline and
  $\frac {1} {R} =\kappa $, the curvature of the streamlines. This formally integrates to:
 
 \begin{align} \label{eq:38}
 -2\int \kappa \hspace{.1cm} d\eta = \ln \left( \frac {v^2} {2} \right) +constant
 \end{align}

The percentage change in the kinetic energy/mass is given by (7.7):

\begin{align} \label{eq:39}
 -2 \kappa \hspace{.1cm} \Delta \eta = \frac {2} {v^2} \hskip.3em \Delta \left( \frac {v^2} {2} \right) 
 \end{align}

\section{\label{sec:level8}Conclusions}

 We analyze the Navier Stokes equation, and show that all irrotational, non-viscous compressible flows are barotropic. As far as we know, this has never been stated in the literature before, and indirectly suggests that vorticity is required to make flows non-barotropic.

Also we derive that the Bernoulli function, $Be=\pa {\Phi} {t}  + \frac {1} {2} \bm\nabla \Phi \cdot \bm\nabla \Phi  $, is only a function of $\rho (\mathbf r, t)$ (even for viscous flow). 
 The ideal gas law then shows that the temperature is a function of $\rho $, and the first law of thermodynamics indicates that the entropy, s,  and enthalpy, h, are functions of $\rho $ (for non-viscous flow). The level surfaces of $ \rho  $, P, Be, T,  s , and h  undulate in time but move as one, although the time and spatial dependence of each variable may be quite different. The irrotational speed is only a function of density for steady, viscous flow. 
 
 We also find that steady flows have a Bernoulli function that is constant in all space, not just along streamlines. 
 This is required for irrotational, compressible, viscous, steady flow. 

\subsection*{Acknowledgment}
We thank E. J. Schmahl for useful discussions. 

\subsection*{Conflicts of Interest}

On behalf of all authors, the corresponding author states that there is no conflict of interest.


\end{document}